\documentclass[preprint,times]{aastex61}

\usepackage{bm}

\received{2017 March 23}
\revised{2017 April 6}
\accepted{2017 April 6}
\submitjournal{the Astrophysical Journal Letters}

\shorttitle{Radiation pressure on interstellar dust}
\shortauthors{Kimura}

\begin{document}

\title{High Radiation Pressure on Interstellar Dust Computed by Light-Scattering Simulation on Fluffy Agglomerates of Magnesium-silicate Grains with Metallic-iron Inclusions}

\correspondingauthor{Hiroshi Kimura}
\email{hiroshi\_kimura@cps-jp.org}

\author{Hiroshi Kimura}
\altaffiliation{Visiting Scientist, Center for Planetary Science (CPS), Chuo-ku Minatojima Minamimachi 7-1-48,
Kobe 650-0047, Japan}
\affiliation{Division of Particle and Astrophysical Science, Graduate School of Science, Nagoya University, Furo-cho, Chikusa-ku, Nagoya 464-8602, Japan}



\begin{abstract}
Recent space missions have provided information on the physical and chemical properties of interstellar grains such as the ratio $\beta$ of radiation pressure to gravity acting on the grains in addition to the composition, structure, and size distribution of the grains.
Numerical simulation on the trajectories of interstellar grains captured by {\it Stardust} and returned to Earth constrained the $\beta$ ratio for the {\it Stardust} samples of interstellar origin.
However, recent accurate calculations of radiation pressure cross-sections for model dust grains have given conflicting stories in the $\beta$ ratio of interstellar grains.
The $\beta$ ratio for model dust grains of so-called ``astronomical silicate'' in the femto-kilogram range lies below unity, in conflict with $\beta \sim 1$ for the {\it Stardust} interstellar grains.
Here, I tackle this conundrum by re-evaluating the $\beta$ ratio of interstellar grains on the assumption that the grains are aggregated particles grown by coagulation and composed of amorphous MgSiO$_{3}$ with the inclusion of metallic iron.
My model is entirely consistent with the depletion and the correlation of major rock-forming elements in the Local Interstellar Cloud surrounding the Sun and the mineralogical identification of interstellar grains in the {\it Stardust} and {\it Cassini} missions.
I find that my model dust particles fulfill the constraints on the $\beta$ ratio derived from not only the {\it Stardust} mission but also the {\it Ulysses} and {\it Cassini} missions.
My results suggest that iron is not incorporated into silicates but exists as metal, contrary to the majority of interstellar dust models available to date.
\end{abstract}




\section{Introduction} \label{sec:intro}

To understand the physical and chemical properties of interstellar dust remains a long-standing problem, since it is common practice to indirectly infer the properties of the dust from astronomical observations, which have been the only tools to access the information in the last century.
Thanks to the motion of the solar system with respect to the Local Interstellar Cloud (LIC), recent space missions mark the beginning of a new era that interstellar grains are directly accessible in the inner solar system \citep{bertaux-blamont1976,mann-kimura2000}.
The first clear evidence for the detection of LIC grains penetrating into the Solar System was provided by the impact ionization detector on board {\it Ulysses} \citep{gruen-et-al1994}.
Subsequently, the identification of LIC grains has been reported by similar detectors on board {\it Hiten}, {\it Galileo}, and {\it Nozomi}, and also confirmed by reanalysis of Helios' data \citep{svedhem-et-al1996,gruen-et-al1997,altobelli-et-al2006,sasaki-et-al2007}.
The {\it Stardust} mission has provided a great opportunity to identify the composition of LIC dust by analyzing genuine samples of the dust in the laboratory \citep{westphal-et-al2014b}.
However, the trajectory of {\it Stardust} did not allow the Stardust Interstellar Dust Collector to capture a sample of LIC grains with $\beta > 1.6$ where $\beta$ is the ratio of solar radiation pressure to solar gravity \citep{westphal-et-al2014a}.
Here, the condition of $\beta > 1.6$ corresponds to the mass range of $1.0\times{10}^{-17} \leqq  m \leqq 3.2\times{10}^{-17}~\mathrm{kg}$, based on the {\it Ulysses} in situ measurements of LIC grains \citep{kimura-et-al2003}.
Numerical simulation of trajectories for the {\it Stardust} samples suggests that the $\beta$--$m$ curve intersects $\beta = 1$ around $m \sim 3 \times{10}^{-15}~\mathrm{kg}$ \citep{sterken-et-al2014}.
Accordingly, \citet{silsbee-draine2016} attempted to model LIC grains that have $\beta = 1$ at $m \sim 3 \times{10}^{-15}~\mathrm{kg}$ using aggregated particles of amorphous silicate grains as well as composite particles of amorphous silicate grains and metallic iron or amorphous carbon grains.
They failed to reproduce $\beta = 1$ at $m \sim 3 \times{10}^{-15}~\mathrm{kg}$ for their plausible dust models, resulting in a question as to whether the {\it Stardust} samples are of interstellar origin.
They computed the $\beta$--$m$ curve with the fixed number of constituent grains (monomers), indicating that the size of their particles is determined by the size of monomers.
In reality, the size of aggregated particles most likely increases with the number of monomers rather than the size of monomers, if the particles grew by coagulation.
It is, therefore, important that the the evaluation of $\beta$--$m$ curve is based on proper assumptions about the structure and composition of dust particles.

Recently, the {\it Cassini} mission has revealed that the chemical composition of LIC dust of mass $m \la 5\times{10}^{-16}~\mathrm{kg}$ is essentially solar with the equal partition of Mg, Si, and Fe \citep{altobelli-et-al2016}.
The ion amplitudes measured by the Chemical Analyzer subsystem of the Cosmic Dust Analyzer (CDA) on board {\it Cassini} indicate that the grains are not fluffy, but compact in the mass range mentioned above. 
It turned out that the $\beta$ values derived from the data analyses of CDA are consistent with compact grains of amorphous MgSiO$_3$ with the inclusion of metallic iron.
Owing to the insensitivity of the CDA to large grains, the result of CDA does not rule out the possibility of aggregated particles for $m \ga 5\times{10}^{-16}~\mathrm{kg}$.
A comparison of {\it Ulysses} in situ data with numerical simulation of dust trajectories suggests that LIC grains with mass $m > 2\times{10}^{-16}~\mathrm{kg}$ are fluffy \citep{sterken-et-al2015}.
All the in situ data seem to be in harmony only if LIC grains of $m \la  2\times{10}^{-16}~\mathrm{kg}$ are compact and LIC grains of $m >  2\times{10}^{-16}~\mathrm{kg}$ are fluffy \citep{kimura-et-al2016}.
I shall verify this hypothesis by demonstrating that the $\beta$--$m$ curve for aggregated particles fulfills the constraint of $\beta = 1$ around $m \sim 3 \times{10}^{-15}~\mathrm{kg}$.

\section{Dust Model} \label{sec:model}

\subsection{The Configuration of Monomers}

I assume that interstellar dust undertakes coagulation growth in dense molecular clouds by the ballistic cluster--cluster agglomeration (BCCA) process.
In my simulation, BCCA particles are built under the coagulation growth that is conducted to hit and stick on respective contacts of two equal aggregated particles consisting of identical spherical grains (monomers) with a radius of $a_\mathrm{m} =0.1~\micron$.
The mass $m$ of an aggregated particle is proportional to the number $N_\mathrm{m}$ of monomers as $m = N_\mathrm{m} (4/3) \pi a_\mathrm{m}^3 \rho_\mathrm{m}$ with $\rho_\mathrm{m}$ being the density of monomers.

\subsection{The Composition of Monomers}

The chemical composition and the $\beta$ ratio of LIC grains measured by CDA/{\it Cassini} were best explained by compact grains composed of amorphous $\mathrm{MgSiO_3}$, in which metallic Fe is embedded, with equal abundances of Mg, Si, and Fe \citep{altobelli-et-al2006}.
Therefore, I may consider that aggregated particles consist of grains composed of amorphous $\mathrm{MgSiO_3}$ with the inclusion of metallic Fe along with the elemental abundance ratio of $\mathrm{Si}:\mathrm{Mg}:\mathrm{Fe} = 1:1:1$ for simplicity.
This elemental abundance ratio results in the average grain density of $\rho_\mathrm{m} = 4.16\times{10}^{3}~\mathrm{kg~m^{-3}}$ on the assumption that the density of amorphous $\mathrm{MgSiO_3}$ and that of metallic Fe are given by $3.30\times{10}^{3}~\mathrm{kg~m^{-3}}$ and $7.86\times{10}^{3}~\mathrm{kg~m^{-3}}$, respectively.

\section{Computational Procedure} \label{sec:method}

\subsection{Dielectric Function}

I apply the Maxwell--Garnett mixing rule to estimate the effective dielectric function $\epsilon(\lambda)$ of amorphous $\mathrm{MgSiO_3}$ with the inclusion of metallic Fe using their volume fractions of 81.07\% and 18.93\%, respectively \citep{garnett1904}.
The dielectric function of amorphous MgSiO$_3$ is taken from \citet{scott-duley1996} for wavelengths up to $\lambda = 62~\micron$ and the values for longer wavelengths are extrapolated on the assumption that the real part is constant and the imaginary part is proportional to $\lambda^{-1}$.
The dielectric function of metallic Fe is taken from \citet{moravec-et-al1976}, \citet{johnson-christy1974}, and \citet{ordal-et-al1988} for wavelengths up to $\lambda = 286~\micron$ and the values for longer wavelengths are linearly extrapolated.

\subsection{Radiation Pressure Cross-section}

I use the superposition {\it T}-matrix method to compute the radiation-pressure cross-section $C_\mathrm{pr}(\epsilon)$ for aggregated particles whose composition is characterized by the effective dielectric function $\epsilon(\lambda)$ \citep{mackowski-mishchenko1996}.
The computation of radiation-pressure cross-section is performed in the wavelength range of $\lambda = 0.14$--$300~\micron$ because the contributions of shorter and longer wavelengths to the radiation pressure are negligibly small.
By taking advantage of the superposition {\it T}-matrix method, I shall calculate the radiation-pressure cross-section averaged over particle orientations.

\subsection{$\beta$ Ratio}

The $\beta$ ratio of dust particles in random orientations is given by the following equation\footnote{In general, the direction of radiation pressure force acting on aggregated particles is not antiparallel to the direction of gravitational force, unless random orientations are considered \citep{kimura-et-al2002a}.} \citep{kimura-et-al2004}:
\begin{eqnarray}
\beta = \frac{\pi R_\sun^2}{G M_\sun m c} \int_{0}^{\infty} B_\sun C_\mathrm{pr} \,d\lambda ,
\end{eqnarray}
where $R_\sun$, $M_\sun$, and $B_\sun$ are the radius, the mass, and the radiance of the Sun, respectively, and $c$ is the speed of light in vacuum.
According to \citet{silsbee-draine2016}, I approximate the solar radiance by the Planck function with an effective temperature of $5777~\mathrm{K}$.
\citet{silsbee-draine2016} used the rectangle method to integrate the product of solar radiance and radiation-pressure cross-section over the wavelength range of $\lambda = 0.27$--$3.34~\micron$ with the fractional accuracy of $(1$--$5)\times{10}^{-3}$.
In contrast, I use the Romberg's method for the integration of the product over the wavelength range of $\lambda = 0.14$--$300~\micron$ with the fractional accuracy of ${10}^{-8}$.

\section{Numerical Results} \label{sec:results}

\begin{figure}[t]
\epsscale{.60}
\plotone{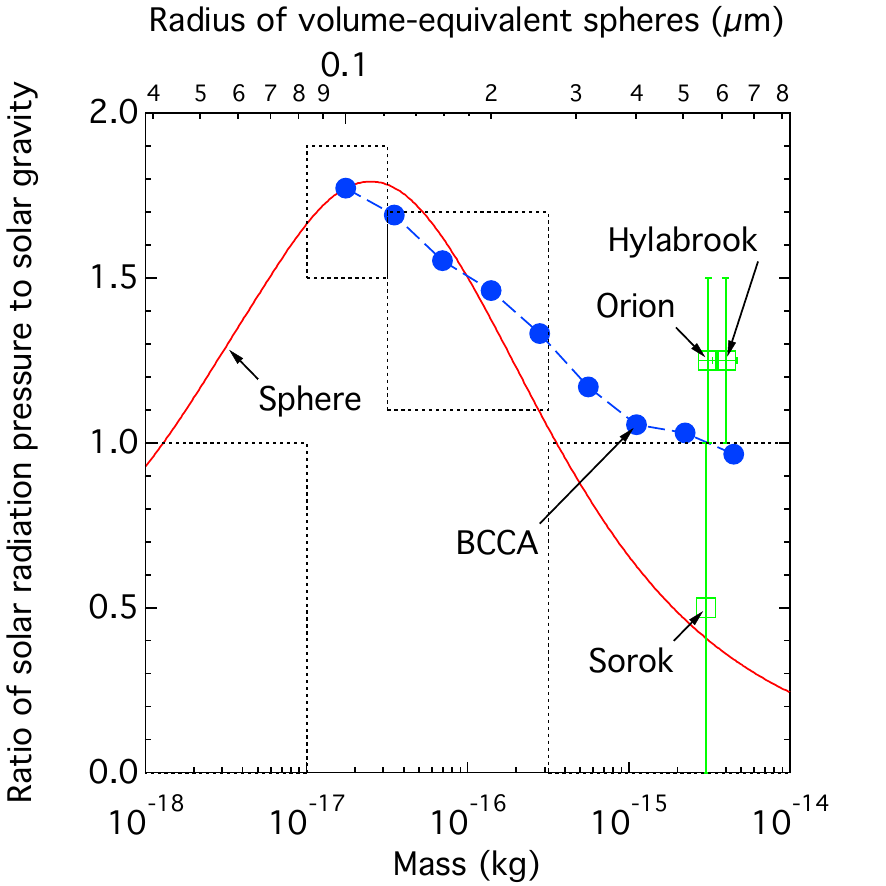}
\caption{$\beta$ ratio of solar radiation pressure to solar gravity acting on aggregated particles (solid circles) and spherical particles (solid line) as a function of the mass $m$ of the particles.
The boxes indicate the constraints of the $\beta$--$m$ curve derived from the data analysis of {\it Ulysses} in situ impact experiments \citep{kimura-et-al2003}.
The open squares with error bars are the $\beta$--$m$ relation for {\it Stardust} samples (Orion, Hylabrook, and Sorok) determined by numerical simulation of their trajectories \citep{sterken-et-al2014}.
\label{fig1}}
\end{figure}
Figure~\ref{fig1} shows my numerical results on the $\beta$ ratio for different sizes of aggregated particles (solid circles with a dashed line) and compact spheres (solid line).
Also shown are the constraints of the $\beta$--$m$ curve derived from the data analysis of {\it Ulysses} in situ impact experiments (boxes) and numerical simulation on the trajectories of Stardust samples (open squares with error bars) \citep[see][]{kimura-et-al2003,sterken-et-al2014}.

The $\beta$ ratio for aggregated particles decreases with mass, but the mass dependence of the $\beta$ ratio for aggregated particles is weaker than that for compact spherical particles.
As a result, the $\beta$--$m$ curve for aggregated particles intersects $\beta = 1$ at heavier particles than that for compact particles of the same composition.
This weak mass dependence of the $\beta$ ratio for aggregated particles is entirely consistent with previous studies on the $\beta$ ratios for aggregated particles of silicates and carbonaceous materials \citep[e.g.,][]{mukai-et-al1992,kimura-et-al2002a,koehler-et-al2007}.
My results indicate that the $\beta$--$m$ curve intersects $\beta = 1$ at $m \approx 3.3 \times{10}^{-15}~\mathrm{kg}$ for aggregate particles and $m \approx 3.3 \times{10}^{-16}~\mathrm{kg}$ for compact particles.

\section{Discussion} \label{sec:discuss}

My results have revealed the similarity of the $\beta$ ratio between aggregated particles and compact particles as far as the mass range of $m <  5\times{10}^{-16}~\mathrm{kg}$ is concerned (see Fig.~\ref{fig1}).
It is, therefore, little wonder that interstellar dust in the mass range of $2\times{10}^{-16} < m < 5\times{10}^{-16}~\mathrm{kg}$ appeared as fluffy grains in the {\it Ulysses} data but as compact grains in the {\it Cassini} data \citep[see][]{sterken-et-al2015,altobelli-et-al2016}.
It is simply that the $\beta$ ratio for interstellar dust alone does not differentiate fluffy grains from compact grains if the mass of the dust lies below $5\times{10}^{-16}~\mathrm{kg}$.
Consequently, I conclude that the fluffy structure of aggregated particles manifests in the $\beta$ ratio only for large grains of $m \ga 5\times{10}^{-16}~\mathrm{kg}$.

The result of {\it Stardust} is consistent with my results on the intersection of the $\beta$--$m$ curve for aggregated particles and $\beta = 1$ at $m \approx 3 \times{10}^{-15}~\mathrm{kg}$.
It has been known that fluffy particles tend to have a higher $\beta $ value than compact particles of the same mass as the size of the particles increases \citep[e.g.,][]{mukai-et-al1992,koehler-et-al2007}.
Nevertheless, \citet{silsbee-draine2016} claimed that aggregated particles of $m \sim 3\times {10}^{-15}~\mathrm{kg}$ do not have $\beta \approx 1$ unless an unrealistic composition is assumed.
In their model, the size of the particles was controlled by the radius $a_\mathrm{m}$ of monomers, while I have enforced the coagulation growth where the size of aggregated particles is determined by the number $N_\mathrm{m}$ of monomers, instead of the radius $a_\mathrm{m}$.
Moreover, they assumed that amorphous silicates and metallic iron are not intermingled, but the results of {\it Stardust} and {\it Cassini} suggest a mixture of amorphous silicates and metallic iron on a nanometer scale as I have assumed here \citep[see][]{westphal-et-al2014b,altobelli-et-al2016}.
Consequently, my success in reproducing $\beta > 1$ for $m \leqq 3\times {10}^{-15}~\mathrm{kg}$ can be attributed to my realistic assumption about the composition and structure of the particles.

To elucidate the origin of the difference in the results between \citet{silsbee-draine2016} and this Letter, I shall compare the $\beta$ values for aggregated particles of $N_\mathrm{m}=32$ grains.
\citet{silsbee-draine2016} constructed aggregated particles of $N_\mathrm{m}=32$ grains using the ballistic particle--cluster agglomeration (BPCA) process, instead of the BCCA process used in this Letter.
While the $\beta$ values in the femto-kilogram range are higher for BCCA particles than BPCA particles, the $\beta$ values at $N=32$ and $a_\mathrm{m} \approx 0.1~\micron$ (i.e., $m \sim 6 \times {10}^{-16}$) are expected to be nearly independent of the coagulation process \citep[cf.][]{koehler-et-al2007}.
Nevertheless, there is a noticeable difference in the results at $N_\mathrm{m}=32$ between \citet{silsbee-draine2016} and this Letter, indicating that the difference rather originates from the contribution of metallic iron to the optical properties of aggregated particles.
My results imply that radiation pressure becomes stronger if iron is imbedded in amorphous magnesium silicates than if the particles contain iron monomers separately from silicate monomers.
Therefore, I conclude that the origin of the difference in the results between \citet{silsbee-draine2016} and this Letter comes from not only the assumption of coagulation process but also the method of incorporating iron inclusions in aggregated particles.

The majority of interstellar dust models assume amorphous silicates of stoichiometrically olivine ($\mathrm{MgFeSiO_4}$) or pyroxene ($\mathrm{MgFeSi_2O_6}$) composition in interstellar dust \citep[e.g.,][]{draine-lee1984,jaeger-et-al1994}.
However, the gas phase of the LIC does not show a correlation between Mg and Fe, but Mg and Si, implying that silicates are magnesium-rich and iron-poor \citep{kimura2015}.
Accordingly, the minority of interstellar dust models assume amorphous silicates of stoichiometrically forsterite ($\mathrm{Mg_2SiO_4}$) or enstatite ($\mathrm{MgSiO_3}$) composition in interstellar dust \citep[e.g.,][]{kimura-et-al2003,jones-et-al2013}.
In general, the $\beta$ ratio for silicate grains does not exceed unity in the Sun's radiation field, irrespective of the composition and structure of the grains \citep{kimura-et-al2002b}.
On the contrary, if magnesium-rich, iron-poor silicate grains contain metallic iron as inclusions, then the maximum $\beta $ value may exceed unity \citep{altobelli-et-al2016}.
This is also the case for fluffy agglomerates consisting of submicron magnesium-silicate grains with metallic iron inclusions as demonstrated in this Letter.
Therefore, the identification of LIC grains with $\beta > 1$ by recent space missions may be an additional indication of iron being inclusions of silicates in a metallic form.

I have shown that the intersection of the $\beta$--$m$ curve and $\beta = 1$ takes place at $m \approx 3\times {10}^{-15}~\mathrm{kg}$ for aggregated particles of $0.1~\micron$ radius monomers.
It is worth noting that the critical mass of aggregated particles at which the $\beta$--$m$ curve intersects $\beta = 1$ depends on the radius of monomers, $a_\mathrm{m}$ \citep{kimura-et-al2002b}.
A slightly larger monomer size, for example, $a_\mathrm{m} = 0.11~\micron$, raises the critical mass, while an extremely large monomer size, for example, $a_\mathrm{m} = 1~\micron$, lowers the critical mass.
As a result, I cannot help but expect that the actual size of monomers in interstellar grains is not far from my assumption of $a_\mathrm{m} = 0.1~\micron$.

\acknowledgments

I would like to thank Daniel Mackowski, Kirk Fuller, and Michael Mishchenko for the availability of the {\sf scsmtm1} code.
The author is indebted to JSPS's Grants-in-Aid for Scientific Research (KAKENHI \#23244027, \#26400230, \#15K05273).

\end{document}